\documentclass[12pt,a4paper]{article}
\usepackage{color,graphicx}

\def\lvec#1{\setbox0=\hbox{$#1$}
    \setbox1=\hbox{$\scriptstyle\leftarrow$}
    #1\kern-\wd0\smash{
    \raise\ht0\hbox{$\raise1pt\hbox{$\scriptstyle\leftarrow$}$}}
    \kern-\wd1\kern\wd0}
\def\rvec#1{\setbox0=\hbox{$#1$}
    \setbox1=\hbox{$\scriptstyle\rightarrow$}
    #1\kern-\wd0\smash{
    \raise\ht0\hbox{$\raise1pt\hbox{$\scriptstyle\rightarrow$}$}}
    \kern-\wd1\kern\wd0}

\def\diracstar#1#2{
    \setbox0=\hbox{$\gamma$}\setbox1=\hbox{$\gamma_{#1}$}
    \gamma_{#1}\kern-\wd1\kern\wd0
    \smash{\raise4.5pt\hbox{$\scriptstyle#2$}}}

\newcommand{\beq}{\begin{equation}}
\newcommand{\eeq}{\end{equation}}
\newcommand{\beqn}{\begin{eqnarray}}
\newcommand{\eeqn}{\end{eqnarray}}
\newcommand{\nn}{\nonumber}

\begin{document}
\begin{titlepage}
\pagestyle{empty}
\date{}
\title{\bf A novel proof of the DFT formula for the interatomic force field of Molecular Dynamics \vspace*{3mm}}

\author{
        S.\ Morante$^{(a)\footnote{morante@roma2.infn.it}}$\, and\, G.C.\ Rossi$^{(a)(b)\footnote{rossig@roma2.infn.it}}$
}

\maketitle
\begin{center}
  $^{(a)}${\small Dipartimento di Fisica, Universit\`a di  Roma
  ``{\it Tor Vergata}''\\ INFN, Sezione di Roma 2}\\
  {\small Via della Ricerca Scientifica - 00133 Roma, Italy}\\
    \vspace{.1cm}
$^{(b)}${\small Centro Fermi - Museo Storico della Fisica e Centro Studi e Ricerche E.\ Fermi\\
Compendio del Viminale, Piazza del Viminale 1 I-00184 Rome, Italy}
\end{center}

\vspace{.5cm}

\abstract{We give a novel and simple proof of the DFT expression for the interatomic force field that drives the motion of atoms in classical Molecular Dynamics, based on the observation that the ground state electronic energy, seen as a functional of the external potential, is the Legendre transform of the Hohenberg--Kohn functional, which in turn is a functional of the electronic density. We show in this way that the so-called Hellmann--Feynman analytical formula, currently used in numerical simulations, actually provides the exact expression of the interatomic force.}

\vspace{1.5cm}
\end{titlepage}

\newpage
     
\section{Introduction} 
\label{sec:INTRO} 

In the Born--Oppenheimer approximation~\cite{BOA} the atomic motion of classical Molecular Dynamics (MD) is governed by the Newton's equations 
\beq
M_I\frac{d^2 \vec R_I}{dt^2}=-\vec \nabla_I \Big{(}E_0(\{\vec R\,\}) +V_{NN}(\{\vec R\,\}) \Big{)}\, ,
\label{AMD}
\eeq
where $\{\vec R\,\}=\{\vec R_I, I=1,2,\ldots, N_A\}$ is the set of atomic coordinates and $E_0(\{\vec R\,\})$ is the energy of the electronic ground state~\footnote{For the sake of the present argument we ignore spin and all the effects related to the existence and the relevance of excited states and electronic state transitions.}. $V_{NN}[\{\vec R\,\}]$ is the pairwise (repulsive) Coulomb interaction potential between atoms, the explicit expression of which we do not need in this discussion. 

The Schr\"odinger equation for the electronic ground state, $\Phi_0$, in the coordinate representation reads 
\beq
{\cal H}_{\rm el} \Phi_0(\{\vec r\,\};\{\vec R\,\})=E_0(\{\vec R\,\})\Phi_0(\{\vec r\,\};\{\vec R\,\})\, ,
\label{ESE}
\eeq
where  $\{\vec r\,\}=\{\vec r_i, i=1,2,\ldots, N_e\}$ is the set of electronic coordinates and 
\beq
{\cal H}_{\rm el} =-\frac{\hbar^2}{2m_e}\sum_{i}\vec \nabla^2_{i}+\sum_{i<j}\frac{e^2}{|\vec r_i-\vec r_j|}+U(\{\vec r\,\};\{\vec R\,\})\, .
\label{HEL}
\eeq
${\cal H}_{\rm el} $ is the sum of the electronic kinetic term, the pairwise (repulsive) Coulomb interaction between electrons and the (attractive) potential among electrons and atoms. The latter is typically of the form 
\beq
U(\{\vec r\,\};\{\vec R\,\})=-\sum_{i,I}\frac{Z_Ie^2}{|\vec R_I-\vec r_i|}\equiv \sum_i u(\vec r_i;\{\vec R\,\})
\label{EXTP}
\eeq
with $Z_I e$ the (positive) electric charge on the $I$-th atom. It is the dependence of $U$ upon the atomic coordinates that is responsible for the parametric dependence of both the electronic ground state energy, $E_0$, and the wave function, $\Phi_0$, on $\{\vec R\,\}$. 

The aim of this note is to provide a novel, rigorous and generally valid proof of the widely used formula 
\beq
-\vec \nabla_{I} E_0(\{\vec R\,\})=-\int d^3 r\, n_u(\vec r\,;\{\vec R\,\})  \vec \nabla_{I} u(\vec r\,;\{\vec R\,\}) 
\label{EXFOR1}
\eeq
for the atomic force field in eq.~(\ref{AMD}). On the r.h.s.\ of eq.~(\ref{EXFOR1}) $u$ is the one-body ``external potential'' (see its implicit definition in eq.~(\ref{EXTP})) and $n_u$ is the ground state density of the electronic system which in terms of the solution of the Schr\"odinger equation~(\ref{ESE}) is given by 
\beq
n_u(\vec r;\{\vec R\,\})=\int \prod_{i=2}^{N_e} \,d^3r_i\, |\Phi_0(\vec r, \vec r_2, \ldots, \vec r_N;\{\vec R\,\})|^2\, .
\label{EDF}
\eeq
We prove directly eq.~(\ref{EXFOR1}) within the DFT formalism from the observation that the ground state electronic energy $E_0$, seen as a functional of the external potential, $u$, is the Legendre transform of the Hohenberg--Kohn functional, $F^{\rm HK}$, with the latter a functional of the electronic density, $n$~\cite{DFT,BDFT,LIEB,ES}.

The more conventional approach is to derive eq.~(\ref{EXFOR1}) as an application of the Hellmann--Feynman theorem (see refs.~\cite{HELM,FE}~\cite{GU,PA} and the Appendix of this work). We refer the reader to the review paper of ref.~\cite{MH} and the book~\cite{DFTP} for a general discussion on the validity and the applicability of the Hellmann--Feynman formula. 

\subsection{The Hohenberg--Kohn theorem}
\label{sec:HKTHEO}

In the mathematical framework where DFT is formulated, the proof of~(\ref{EXFOR1}) can be directly based on the Hohenberg--Kohn (HK) Theorem. In their seminal paper~\cite{DFT}  HK established the following simple, but far reaching proposition.

\vspace{.2cm}
{\bf HK Theorem -} There exists a bijective correspondence between the elements of the space of ``external potentials'', $u\in {\cal V}$, the elements of the space of electronic densities, $n\in {\cal N}$, and the elements of the space of ``ground state wave functions'', $\Phi_0\in \Gamma$
\beq
{\cal V} \leftrightarrow {\cal N} \leftrightarrow \Gamma\, .
\label{VNP}
\eeq
In simple words, given the ``external potential'' $u$, not only the ground state electronic wave function solution of the Schr\"odinger equation~(\ref{ESE}) and the associated electronic density~(\ref{EDF}) are obviously~\footnote{As  commonly done, we assume here that the solution of~(\ref{ESE}) is non-degenerate.} uniquely determined, but also the converse is true. 
\vspace{.2cm}

Naturally the difficult step is to determine $\Phi_0$, or equivalently $n_u$, once $u$ is given. To this end the most convenient way of formalizing the problem is to make recourse to the basic variational principle of Quantum Mechanics. As is well known, the powerful HK Theorem can be used to reduce the task of solving the multi-body Schr\"odinger equation~(\ref{ESE}) to that of solving the set of one-body Kohn--Sham equations~\cite{KS} at the price of introducing the exchange-correlation functional. 

For the purpose of this note, we need to dwell neither on the well-known proof of this statement nor on the subtle problems concerning the mathematical foundation of the functional variational  principles in DFT~\footnote{The mathematics of the functional Legendre transform in DFT has been beautifully discussed and clarified in refs.~\cite{LIEB,ES}.}. For us it is enough to recall that the formulation of the variational principle in the DFT framework is based on introducing the HK functional of the electronic density 
\beq
F^{\rm HK}[n]=\langle \Phi_0[n]|T+W| \Phi_0[n] \rangle\, ,\qquad n\in {\cal N}\, ,\label{HKUF}
\eeq
where with obvious notations 
\beqn
T= -\frac{\hbar^2}{2m_e}\sum_{i}\vec \nabla^2_{i}\, ,\qquad W = \sum_{i<j}\frac{e^2}{|\vec r_i-\vec r_j|}\, .\label{W}
\eeqn
Since, according to the HK Theorem, there is a one-to-one correspondence between densities ($n\in {\cal N}$) and (ground state) wave functions ($\Phi_0\in \Gamma$), $F^{\rm HK}[n]$ is a well defined and universal functional of $n\in{\cal N}$. Moreover, it can be proved~\cite{KS} that the functional   
\beq
E[n]\equiv F^{\rm HK}[n]+\int d^3r\, u(\vec r\,;\{\vec R\,\}) n(\vec r\,)\, .\label{HKUFN}
\eeq
is stationary at $n=n_u$ (defined in eq.~(\ref{EDF})). Thus the variational principle that, given $u$, determines the actual (ground state) electronic density takes the form~\footnote{The mathematical circumstances under which the stationarity condition for the functional~(\ref{HKUFN}) leads to eq.~(\ref{VP}) have been carefully spelled out in refs.~\cite{LIEB,ES}.} 
\beq
\frac{\delta F^{\rm HK}[n]}{\delta n(\vec r\,)} +u(\vec r\,;\{\vec R\,\})=0\, .
\label{VP}
\eeq
The solution of this equation (which is equivalent to the Schr\"odinger equation~(\ref{ESE})) provides the desired electronic density, that is to say, the value of $n$ that is related to $u$ under the correspondence~(\ref{VNP}). 

The quantity of interest for computing the force in eq.~(\ref{AMD}) is the expectation value of ${\cal H}_{\rm el}$ in the ground state solution of the Schr\"odinger equation~(\ref{HEL}), i.e.\ the value of $E[n]$ at its stationary point, that in terms of $F^{\rm HK}$ is given by 
\beqn
\hspace{-1.4cm}&&E_0(\{\vec R\,\})=\langle \Phi_0[n]| {\cal H}_{\rm el} |\Phi_0[n] \rangle\Big{|}_{n=n_u}=\nn\\
\hspace{-1.4cm}&&\quad=F^{\rm HK}[n_u]+\int d^3 r\, u(\vec r\,;\{\vec R\,\}) n_u(\vec r\,;\{\vec R\,\})\equiv E_0[u]\, .
\label{EXVH}
\eeqn

\section{Computing the atomic force $-\vec \nabla_{I} E_0(\{\vec R\,\})$}
\label{COFO}

We now give a proof of~(\ref{EXFOR1}) directly within the DFT formalism. The argument goes as follows. 

One can look at the mathematical steps leading from the HK functional (eq.~(\ref{HKUF})) to the ground state energy functional (eq.~(\ref{EXVH})) as the steps that one needs to  follow to pass from $F^{\rm HK}[n]$ to its the Legendre transform. The latter can be, in fact, constructed from the defining formula    
\beq
{\cal L}\left(F^{\rm HK}\right)\![u] = {\mbox{inf}}_{\,n\,({\rm fixed} \,u)} \Big{[}F^{\rm HK}[n]+\int d^3 r \, u(\vec r\,;\{\vec R\,\}) n(\vec r\,)\Big{]}\, .
\label{LETR}
\eeq
Under sufficiently general assumptions valid for most of the interesting applications~\cite{LIEB,ES}, the stationarity condition ensuing from~(\ref{LETR}) is precisely eq.~(\ref{VP}). By evaluating the functional within the square brackets of (\ref{LETR}) at its stationary point, $n_u$, one immediately recognizes that ${\cal L}(F^{\rm HK})[u]$ coincides with the expression in the second line of eq.~(\ref{EXVH}), i.e.\ with the ground state electronic energy 
\beq
{\cal L}(F^{\rm HK})[u]=E_0[u]\, .
\label{EQUA}
\eeq
Consistently with the general HK theorem (that establishes the existence of the one-to-one correspondence~(\ref{VNP})), this equality explicitly shows that the electronic ground state energy, $E_0$, can be regarded as a universal functional of the external potential. Eq.~(\ref{EQUA}) also implies the obvious, but important for us, fact that the dependence of $E_0$ on the atomic coordinates is only through the external potential, $u$. 

By standard manipulations one can compute the functional derivative of $E_0[u]$ with respect to $u$, getting for the Legendre dual of eq.~(\ref{VP})
\beqn
\frac{\delta E_0[u]}{\delta u(\vec r\,)} = \int d^3 r{'} \, \frac{\delta F^{\rm HK}[n]}{\delta n(\vec r\,')}\Big{|}_{n=n_u} \frac{\delta n_u(\vec r\,')}{\delta u(\vec r\,)}+\int d^3r' u(\vec r\,')\frac{\delta n_u(\vec r\,')}{\delta u(\vec r\,)}+n_u(\vec r\,)\, .
\label{DVP}
\eeqn
For simplicity of notations in eq.~(\ref{DVP}) we have omitted the parametric dependence on the atomic coordinates. Reinstating it and noting that the first two terms on the r.h.s.\ of eq.~(\ref{DVP}) exactly cancel, owing to the stationarity condition~(\ref{VP}), one obtains
\beqn
\frac{\delta E_0[u]}{\delta u(\vec r\,)}\Big{|}_{u=u(\vec r\,;\{\vec R\,\})}=n_u(\vec r\,;\{\vec R\,\})\, .
\label{DVPF}
\eeqn
This formula immediately leads to eq.~(\ref{EXFOR1}) that follows from the derivative chain 
\beqn
\hspace{-1.cm}&&-\vec \nabla_{I} E_0(\{\vec R\,\})=- \int d^3 r\, \frac{\delta E_0[u]}{\delta u(\vec r\,)}\Big{|}_{u=u(\vec r\,;\{\vec R\,\})} \vec  \nabla_{I} u(\vec r\,;\{\vec R\,\}) =\nn\\
\hspace{-1.cm}&&\qquad\qquad\qquad\,\,= - \int d^3 r\, n_u(\vec r\,;\{\vec R\,\}) \vec \nabla_{I} u(\vec r\,;\{\vec R\,\})  \, .
\label{EXFIN}
\eeqn
We close our discussion with two important observations. 

The first concerns the role of Pulay forces~\cite{PULAY} in the computation of atomic forces. The derivation of eq.~(\ref{EXFOR1}) that we have presented in this work shows that, unlike what it looks in the standard Hellmann--Feynman proof (that we recall in the Appendix), no Pulay forces can ever come into play. The reason why they do not show up in the functional formalism is that, as we have seen, in the algebraic manipulations leading to~(\ref{EXFIN}), $E_0$ is regarded as a functional of the external potential. Thus nowhere the $\vec R$-gradient of the wave-function, which instead enters the expression of the Pulay forces, needs to be evaluated. The consequence of this remark is that the lack of accuracy that might affect the evaluation of the atomic forces provided by eq.~(\ref{EXFOR1}) should not be blamed to unaccounted for Pulay forces~\cite{PULAY} but to a possibly inaccurate solution of the variational problem leading to a bad electronic density, $n_u$.

The second concerns eq.~(\ref{DVPF}). We remark that, although conceptually at the basis of the ``potential functional theory'' (PFT), just like its Legendre dual~(\ref{VP}) is at the basis of the ``density functional theory'' (DFT), eq.~(\ref{DVPF}), at least in its present form, was not noticed in the refs.~\cite{CLEGB,CGB,BW,YAW} where PTF was introduced and developed.

\section{Conclusions}
\label{sec:CONCL}

In this note we have given an elegant proof of eq.~(\ref{EXFOR1}) within the functional formalism using the notion of Legendre transform. The line of arguments we have developed represents an improvement in two respects over to the standard approach based on the Hellmann--Feynman theorem (HFT) recalled in the Appendix. 

First of all the proof we have given sheds a new light on the role of the Pulay correction in the calculation of MD atomic forces. As is well known, in the standard proof of the HFT (see Appendix) the Pulay forces, that come about when the trial wave-function is not an exact eigenstate of the Hamiltonian or the basis set is incomplete, depend on the $\vec R$-gradient of the trial wave-function. In the derivation of eq.~(\ref{EXFOR1}) we give in this paper nowhere the $\vec R$-gradient of the wave-function needs to be computed. In fact, in the derivative chain leading to eq.~(\ref{EXFIN}) the $\vec R$-gradient of the $E_0$ functional is calculated via the dependence of the latter upon the external potential. We are thus led to conclude that the Pulay forces are a fake of the traditional strategy used to prove the HFT which does not take into account the peculiarities of the specific variational problem at hand. 

Naturally the issue of the basis set completeness does matter when the variational equations for $n_u$ are being solved, but it is immaterial as for the question of whether the second line of eq.~(\ref{DERIV1}) vanishes or not. As a consequence ``correcting'' the Hellmann--Feynman formula by taking into account Pulay forces does not necessarily make atomic forces calculation more accurate. It might very well make it even worse. 

Secondly, framing the problem of computing MD atomic forces in the functional formalism has allowed us to ``discover'' eq.~(\ref{DVPF}) which, although at the basis of the PFT, it was not noticed as such in the refs.~\cite{CLEGB,CGB,BW,YAW} where PFT was first discussed.

\vspace{.3cm}

{\bf Acknowledgements - } We wish to thank the anonymous referee for her/his very useful observations that made us to rethink the presentation of the whole paper.
 
\appendix 
\renewcommand{\thesection}{A} 
\section{The standard proof of the Hellmann--Feynman theorem}  
\label{sec:APPA} 

We report here for completeness the standard proof of the Hellmann--Feynman theorem (HFT) and we recall the difficulties of its numerical implementation discussed in the literature (see for instance the review paper~\cite{MH} and ref.~\cite{VP}). We also remind the reader the canonical solutions proposed in~\cite{HUR,PULAY} to circumvent these problems and we contrast them with the clear-cut solution entailed by the results of this paper.  

Let ${\cal H}({R})$ be a  quantum Hamiltonian (i.e.\ a self-adjoint operator in its definition domain~\cite{EFGC,CARFI}) that smoothly depends on the parameter $R$ and let $|\psi_k(R)\rangle$ be its normalized eigenvectors such that (we use the Dirac formalism and to simplify notations we label energy eigenvectors with a discrete index)
\beq
{\cal H}(R)|\psi_k(R)\rangle= E_k(R) |\psi_k(R)\rangle\, .
\label{HEIG}
\eeq
Consider the functional 
\beq
{E}(\psi,R)= \frac{\langle \psi|{\cal H}(R)|\psi\rangle}{\langle\psi|\psi\rangle}\, .
\label{FUNC1}
\eeq
${E}(\psi,R)$ is stationary in correspondence to each eigenstate of the Hamiltonian where obviously one has $E(\psi_k,R)=E_k(R)$. The HFT 
\beqn
\hspace{-1.4cm}&&\frac{\partial {E}_k(R)}{\partial R}= \frac{1}{\langle\psi_k|\psi_k\rangle}\langle \psi_k|\frac{\partial {\cal H}(R)}{\partial R}|\psi_k\rangle
\label {HFT}
\eeqn
follows by taking the derivative of eq.~(\ref{FUNC1}) with respect to $R$ and then setting $|\psi\rangle=|\psi_k\rangle$. With a little algebra one gets, in fact    
\beqn
\hspace{-1.4cm}&&\frac{\partial {E}(\psi,R)}{\partial R}= \frac{1}{\langle\psi|\psi\rangle}\Big{[}\langle \psi|\frac{\partial {\cal H}(R)}{\partial R}|\psi\rangle+\nn\\
\hspace{-1.4cm}&&+\langle \frac{\partial \psi}{\partial R}|{\cal H}(R)-E(\psi,R)|\psi\rangle+
\langle \psi|{\cal H}(R)-E(\psi,R)|\frac{\partial \psi}{\partial R}\rangle\Big{]}\, .
\label {DERIV1}
\eeqn
If $|\psi\rangle=|\psi_k\rangle$ is one of the eigenstates of the Hamiltonian, the terms in the second line of eq.~(\ref{DERIV1}) vanish and one obtains eq.~(\ref{HFT}). 

The problem with the practical implementation of the previous formulae is that in the actual variational calculations one can construct (without much numerical effort) trials vectors, $|\psi_{trial}\rangle$, that satisfy the weaker condition
\beq
\langle\psi_{trial}|{\cal H} -E_{trial}|\psi_{trial}\rangle=0\, ,
\label{TRIALM}
\eeq
rather than 
\beq
{\cal H} |\psi_{trial}\rangle=E_{trial}|\psi_{trial}\rangle\, .
\label{TRIAL}
\eeq
Since $|\psi_{trial}\rangle$ is not guaranteed to be an eigenstate of the Hamiltonian, the terms in the second line of eq.~(\ref{DERIV1}) do not vanish. Two remedies have been proposed in the literature (we follow here the discussion of ref.~\cite{VP}) to get around this problem.

1) One chooses a basis set for $|\psi_{trial}\rangle$ so that the second term in eq.~(\ref{DERIV1}) vanishes (Hurley's condition~\cite{HUR}). One way to achieve this is to take basis sets that do not depend on $R$ (or to have the derivatives of the basis wave-functions with respect to $R$ to be themselves part of the basis set). Plane waves are examples of a basis set belonging to such a category. 

2) One works with the full eq.~(\ref{DERIV1}). The interpretation of this equation is, however, the following. The l.h.s.\ represents the atomic force one wishes to evaluate, while the first term on the r.h.s.\ (which is ``easy'' to compute) is the Hellmann--Feynman force. The last term, referred to as the Pulay~\cite{PULAY} force in the literature, is taken as a correction to the Hellmann--Feynman force. This is the situation one encounters when atom-centered basis wave-functions are used. 

As we have shown in the main text, the actual value of the terms in the second line of eq.~(\ref{DERIV1}) is actually immaterial because eq.~(\ref{EXFOR1}) itself directly follows from the functional variational principle that is at the basis of DFT.

\end{document}